\documentclass[10pt, aps, pra, twocolumn, 
unsortedaddress, superscriptaddress, nofootinbib]{revtex4-2}
\usepackage{graphicx} 
\usepackage{physics}
\usepackage{amsmath}
\usepackage{color}
\usepackage{comment}

\begin{document}
\title{Transformation of vector modes by the Faraday effect in strong magnetic fields}
\author{Sphinx J. Svensson} 
\affiliation{School of Physics and Astronomy, University of Glasgow, Kelvin Building, Glasgow G12 8QQ, United Kingdom}

\author{Craig J. A. Millar}
\affiliation{School of Physics and Astronomy, University of Glasgow, Kelvin Building, Glasgow G12 8QQ, United Kingdom}

\author{Danielle Pizzey}
\affiliation{Department of Physics, University of Durham, Rochester Building, South Road, Durham DH1 3LE, United Kingdom}

\author{Ifan G. Hughes}
\affiliation{Department of Physics, University of Durham, Rochester Building, South Road, Durham DH1 3LE, United Kingdom}

\author{Sonja Franke-Arnold}
\email{Sonja.Franke-Arnold@glasgow.ac.uk} 
\affiliation{School of Physics and Astronomy, University of Glasgow, Kelvin Building, Glasgow G12 8QQ, United Kingdom}

\date{July 2026}

\maketitle
\section{Abstract}
Large Faraday rotations can be generated by circular birefringence of atomic samples in an axial magnetic field in the vicinity of atomic resonance lines. The Faraday angle is a function of the magnetic field strength, the optical density of the atomic sample which may be varied by changing the temperature of the atomic gas, and of course the optical detuning from the transition frequencies. More generally, magneto-optical effects in atomic samples include circular dichroism in addition to birefringence, resulting in a modification of the ellipticity as well as the polarisation alignment. Usually such effects are investigated for homogeneous linear polarisations, but the mechanisms apply also to polarisation structures such as vector vortices. 
We investigate the effect of optical activity of a rubidium vapour in the Hyperfine Paschen-Back regime, for the example of an azimuthally polarised input light beam. We show that for low atomic densities, circular birefringence dominates over dichroism, and azimuthal polarisation is rotated towards radial polarisation. The rotation angle increases with increasing optical densities. At high vapour temperatures, dichroism becomes more and more relevant, leading to intricate variations of both alignment and ellipticity.

\section{Introduction}

Vector light is an umbrella term for light modes which carry spatially dependent polarisation \cite{Rubinsztein-Dunlop/JO:2017,Wang2021}. This feature makes it  attractive for the investigation of light-matter interaction \cite{Wang/AVSQS:2020}, as it enables probing for multiple or all polarisations in a single experiment. 
This new possibility allows for measurement \cite{Fatemi:11,concurrence}, transformation \cite{PhysRevLett.97.113601}, as well as storing \cite{jinwen-skyrmion} of vector light, and opens up new opportunities in quantum sensing \cite{Castellucci/PRL:2021,Ramakrishna-2024}.
Optically active media usually interact with light in a combination of dispersion and absorption. The narrow linewidths of atomic vapours and their easy manipulation via external electromagnetic fields makes them well suited for the experimental study of these phenomena \cite{Babiker_2019}. One of the most famous examples of optical activity is Faraday rotation, i.e. the rotation of linear polarisation in a Faraday medium. 

In atomic vapours, the Faraday effect is caused by circular birefringence induced by Zeeman shifts: In the presence of an external magnetic field, previously degenerate sublevels are shifted in proportion to the magnetic field, as well as the state's associated angular momentum \cite{Steck2001}. The circular birefringence manifests as a rotation of linear polarisation \cite{faraday1846experimental}. This effect is quite famously used in optical isolators \cite{faraday-magnets}, but also has applications in optical switching \cite{optical-switch-ifan, optical-switch-dawes}, far off-resonance laser locking \cite{faraday-stabilise}, magnetometry \cite{budker2000sensitive}, as well as constructing dichroic beamsplitters \cite{faraday-dichroic}. It has also recently been shown to induce spatial rotation in vector light \cite{shiMagneticfieldinducedRotationLight2015, richard-arxiv}.

With strong magnetic fields, the Zeeman shift exceeds the hyperfine interaction, leading to a regime change: The nuclear spin decouples from the electron, and instead both individually couple to the external magnetic field \cite{Sargsyan2015}. The ``good" quantum numbers are no longer $F$ and $m_F$ but instead $m_I$ and $m_J$. In Rb, the hyperfine Paschen-Back (HPB) regime is realised in magnetic field stronger than 0.24 T \cite{ifan-hpb}. The regrouping of energy levels, as shown in figure \ref{fig:theory}, leads to a drastic change in spectrum: Individual hyperfine transitions are GHz detuned from each other, clustered in groups of four. Each individual absorption dip of the cluster comes from a different alignment of the nuclear spin, which is 3/2 for Rb-87. The clusters are effectively sorted by transition type: $\sigma_-$ transitions are red shifted and $\sigma_+$ transitions blue shifted, while the $\pi$ transitions remain relatively close to the original resonance. If the magnetic field is aligned with the optical axis, the $\pi$ transition is usually considered unavailable, though we have recently shown that it becomes accessible in the strong focusing regime \cite{svensson2025}. 

Rotation of radially polarised light was recently reported for a Terbium Gallium Garnet crystal in the Faraday configuration \cite{TAMBAG2023129649} and in an optically active D-fructose solution \cite{xie2026dynamicstransversespinlongitudinal}, showing that the linear polarisation rotation can be observed as a change of the spatial modes. 
In the context of the HPB regime, the Faraday effect was first investigated  by M. Zentile et al \cite{Zentile2014}, where it was demonstrated that the dispersion curves of each absorption dip interfere, as well as that the effect is highly temperature dependent. In this paper, we wish to build on both of these works, demonstrating Faraday rotation of vector light in the HPB regime. 

--- 

\section{theory}
Propagation of light through the atomic vapour, be it homogeneously polarised or vector light, is determined by the frequency dependent complex refractive index $\tilde n=n+i\kappa = \sqrt{1+\chi}$ for the relevant atomic transitions, where $\kappa$ denotes the amplitude absorption coefficient and $n$ is the refractive index. 
In the Faraday configuration, with an external magnetic field along the propagation axis, $\sigma_+$ transitions (corresponding to an increase of the magnetic quantum number by $\Delta m =1$) are driven by left circularly polarised light, whereas $\sigma_-$ transition ($\Delta m =-1$) are driven by right circularly polarised light. 
In an external magnetic field the frequency response of the refractive indices $\tilde n_\pm$ differ for $\sigma_\pm$ transitions, as the absorption and dispersion lines are separated in frequency due to their Zeeman-shift. In atomic vapours, this is the cause of the magneto-optical effects described as circular dichroism and circular birefringence respectively. The former changes the optical ellipticity and is strongest close to resonance, while the latter persists even at large detunings and modifies the polarisation alignment, observed as Faraday rotation. Any linear polarisation is an equal superposition of opposite circular polarisation components and experiences composite absorption and dispersion of the $\sigma_\pm$ transitions. This results in a Faraday angle of
\begin{equation} \label{eq Faraday theory}
\theta_{\rm F} = \frac{2\pi}{\lambda} \int_{-L/2}^{L/2} dz (n_--n_+)/2,
\end{equation}
where $L$ is the length of the vapour cell and $\lambda$ is the wave length.

\begin{figure}
    \centering
    \includegraphics[width=.9\linewidth]{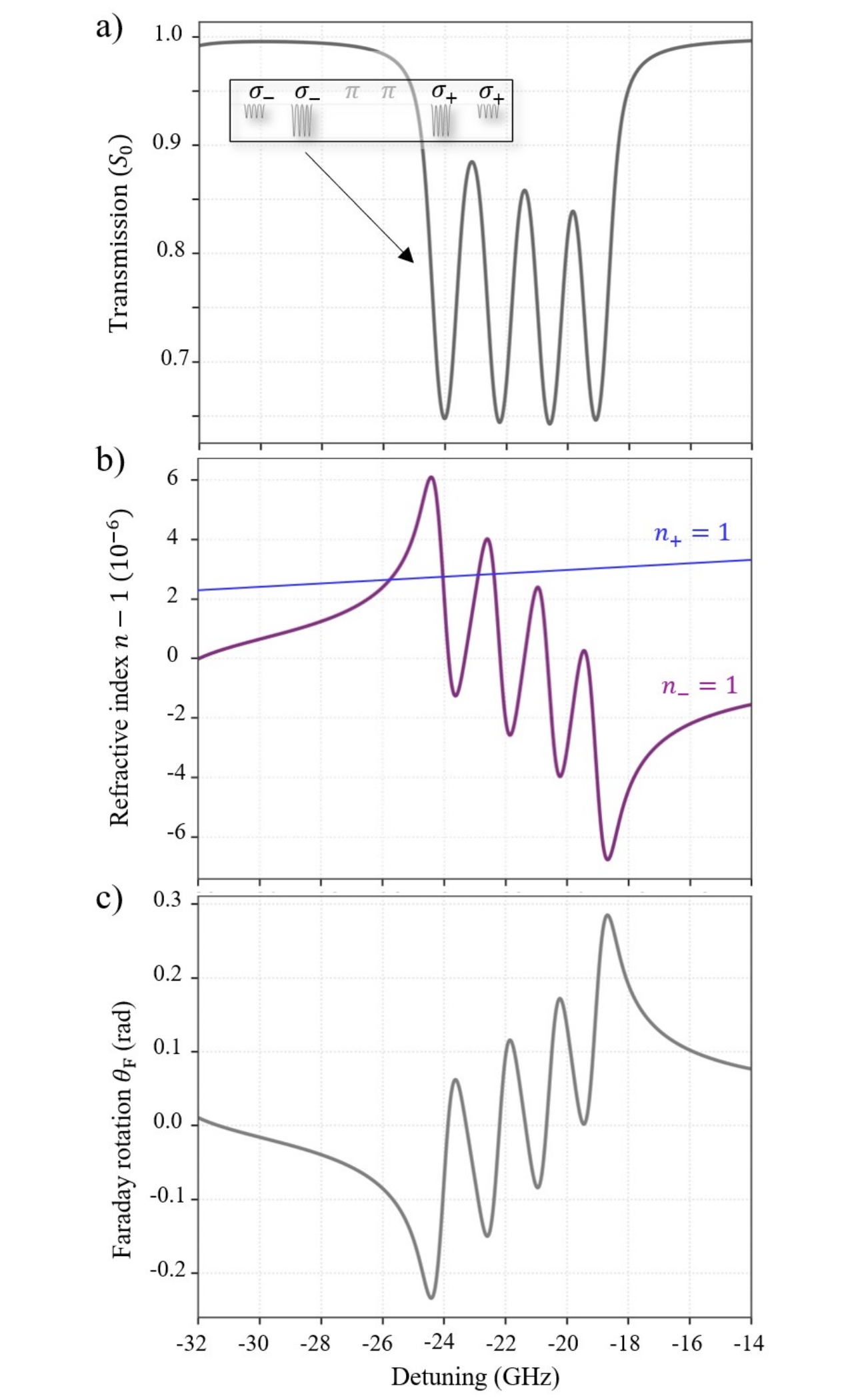}
    \caption{Rb$^{87}$ spectra at $1.6\,\rm{T}$ in the Faraday configuration at 92.1$^\circ$C over the investigated frequency range. The transmission is shown in (a), with the full spectrum from -45 to 45 GHz displayed as inset. The associated refractive indices and the resulting predicted Faraday rotation angle are shown in (b) and (c) respectively. The spectra were generated with the computational modelling tool {\it ElecSus} \cite{Zentile2014}.}
    \label{fig:theory}
\end{figure}

In the HPB regime, where the Zeeman splitting is stronger than the hyperfine interaction, the shift experienced by each energy level is described by the following equation: \cite{Zentile2014}
\begin{equation}
\label{eq:shift}
    \Delta E \approx \left(g_{\rm J} m_{\rm J} \mu_{\rm B} + g_{\rm I} m_{\rm I} \mu_{\rm N} \right)B
\end{equation}
where $B$ is the external magnetic field along the quantisation axis, $g_{\rm I}$ and $g_{\rm J}$ are the gyromagnetic ratios (or Land\'e g factors) of the nucleus and electron respectively, $\mu_{\rm B}$ is the Bohr magneton, and $\mu_{\rm N}$ is the nuclear magneton.

The $\sigma_+$ and $\sigma_-$ transitions are clustered in blocks of four, as shown in the inset of Fig.~\ref{fig:theory} (a), corresponding to the four possible values of $m_{\rm I}$, and spectrally resolved for our magnetic field of $1.6\,\rm{T}$. In this work we investigate the cluster of $\sigma_+$ transitions from the $5S_{1/2}\ket{m_J=-1/2}$ ground states, but similar results hold for all $\sigma_\pm$ transitions.

For the magnetic field of 1.6T used in our experiment, the resonance frequencies for the $\sigma_\pm$ transitions are separated by many times the linewidth. In the frequency range under investigation, the effect of the remote $\sigma_+$ transitions is therefore largely negligible, with $\kappa_+\approx 0$ and $n_+ \approx 0$ and relatively constant across the investigated frequency range (see Fig.~\ref{fig:theory}(b)). On resonance for each of the 4 lines within the $\sigma_-$ transition cluster we therefore expect an increase in ellipticity for the probe light, while the dispersion of the right hand polarised polarisation components should lead to a Faraday rotation, as modelled in Fig.~\ref{fig:theory}(c). The magnitude of this effect depends on the optical density of the atomic vapour which can be increased by increasing the temperature of the atomic vapour.

In order to highlight these magneto-optical effects for polarisation vortices, we investigate vector beams created as equal superpositions of circularly polarised Laguerre-Gauss (LG) modes, and hence featuring linear polarisations that vary across the azimuthal angle $\varphi$:

\begin{align}
    \label{eq:beam}
    \vec{E}_{\rho} & \propto \frac{1}{\sqrt{2}}\left({\rm LG}_0^{1}\hat{\sigma}_-+ {\rm LG}_0^{-1}\hat{\sigma}_+\right) \propto \, \cos \varphi \,\hat{x} + \sin \varphi \,\hat{y} \\ 
    \vec{E}_{\varphi} &\propto \frac{1}{\sqrt{2}}\left({\rm LG}_0^{1}\hat{\sigma}_-- {\rm LG}_0^{-1}\hat{\sigma}_+\right)  \propto \,\sin \varphi \,\hat{x} -\cos \varphi \,\hat{y}.
\end{align}
Here, $\vec{E}_{\rho}$ and $\vec{E}_{\varphi}$ denote the electric field of radially and azimuthally polarised light, and ${\rm LG}_0^{\ell}$ is a Laguerre-Gauss mode with radial quantum number 0 and azimuthal quantum number $\ell$, here taking the values $\pm 1$.  We note, that by changing the relative phase of the circular polarisation components, it is possible to transform a radial beam into an azimuthal beam and vice versa.

Magneto-optical effects apply to the local polarisation states within the vector light profiles, with circular dichroism leading to an overall increase of ellipticity, while circular birefringence, i.e. Faraday rotation, `twists' azimuthal input polarisation towards radial polarisation. 

The described theoretical considerations apply to paraxial light. In our experiment we work with moderately focused light, with a numerical aperture of ${\rm NA} = 0.4.$  In this regime, any radial polarisation components are converted into a superposition of radial and axial light, therefore allowing access to the (strongly off-resonant) $\pi$ transitions as recently demonstrated in \cite{svensson2025}. To limit effects of these $\pi$ transitions we have chosen an azimuthally polarised probe beam, which upon focusing remains azimuthal. We note however, that as Faraday rotation will turn azimuthal light towards radial light, $\pi$ transitions come into play upon propagation -- an effect we intend to study in future work. 

\section{experimental setup}

\begin{figure}
    \centering    \includegraphics[width=\linewidth]{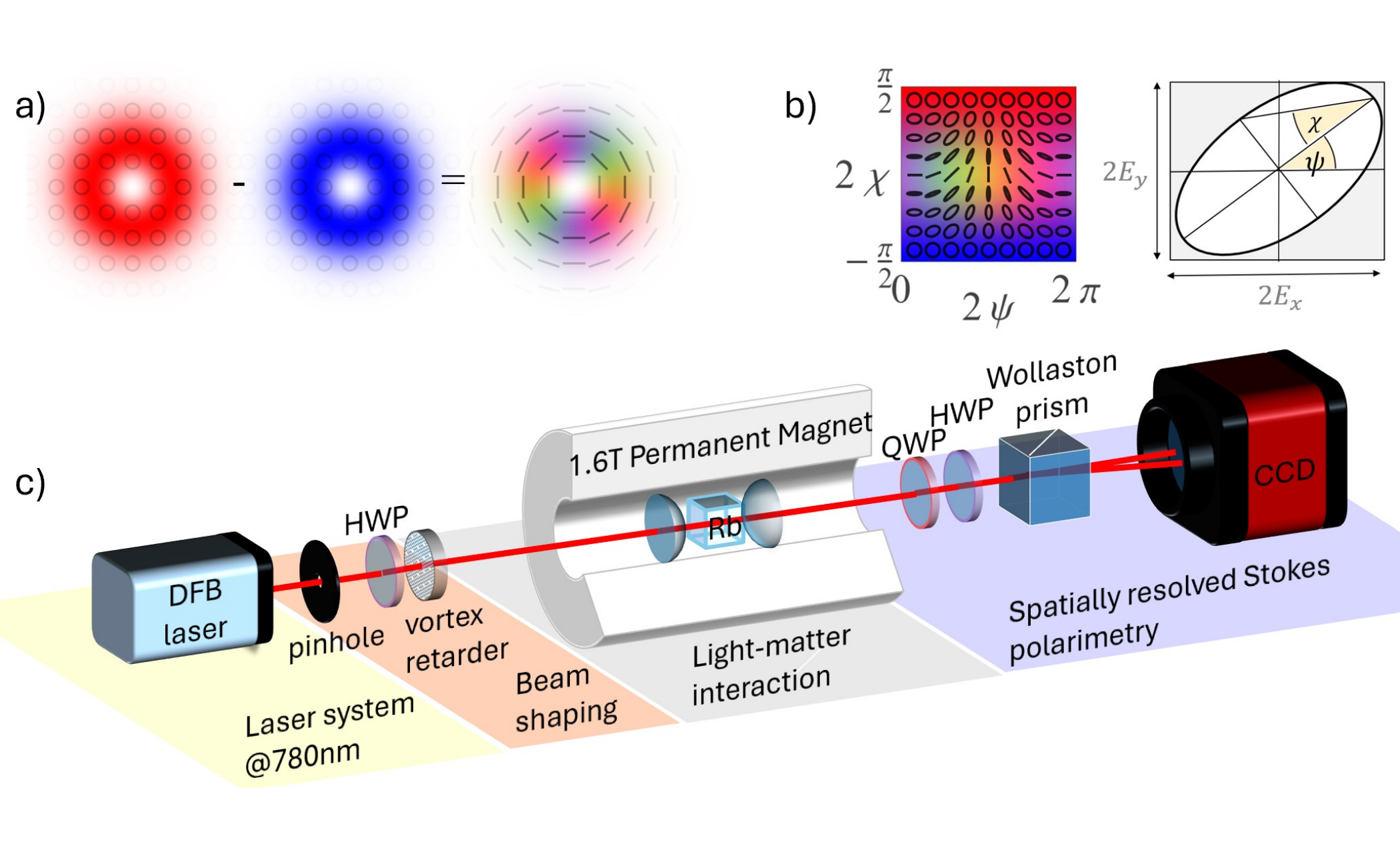}
    \caption{a) Azimuthally polarised probe beam and its decomposition in right and left handed polarisation components. b) Explanation of the colour scheme used to represent polarisation states, parametrised by the eccentricity angle $\chi$ and the alignment angle $\psi$. c) Schematic of the experimental setup.}
    \label{fig:setup}
\end{figure}

In order to measure the effects of near resonant Faraday rotation on vector beams in the HPB regime, the following setup was used: 
A DFB laser scans (mode-hop free) across 50 GHz of detuning, producing a beam that is first spatially filtered with a pinhole and then shaped into an azimuthally polarised vector beam using a m=1 vortex retarder \cite{Marrucci2006}. It is then focused with an NA of 0.4 into a 1mm$^3$ cell containing a Rb-87 vapour. This cell is located in a homogeneous 1.6T magnetic field along the optical axis, produced by a NdFeB permanent magnet~\cite{trenec2011permanent}. 
After interacting with the atoms, the beam is collimated and imaged onto a CCD. Spatially resolved Stokes tomography allows us to reconstruct the polarisation profile for a given frequency. Approximately 250 images are taken over the duration of each (one-way) scan at a frame rate of 50Hz. The measurements are repeated for a range of temperatures between 79$^\circ$C and 132$^\circ$C to investigate the dependence of the effect on number density. These temperatures are estimated by fitting the absorption spectra using {\it Elecsus} \cite{elecsus}. As {\it Elecsus} is strictly valid only for the weak probe regime, neglecting optical saturation and associated bleached absorption, it underestimates the true temperature and atomic density. 

The Faraday rotation angle (as well as the transmission) is then obtained from performing Stokes polarimetry based on overcomplete measurements in 6 polarisation states (horizontal, vertical, diagonal, antidiagonal, right and left). From the Stokes parameters we obtain the orientation of the linear polarisation as 
\begin{equation} \label{eq Faraday experiment}
\theta = \arctan\frac{S_2}{S_1}
\end{equation}
where $S_0$ and $S_1$ denote the total intensity, and the intensity difference in horizontal vs vertical polarisation, and $S_2$ the intensity difference in diagonal and antidiagonal. The Faraday rotation angle is then given by $\theta_{\rm F}=\theta_0-\theta$, where $\theta_0$ is the local polarisation alignment of the input beam. We report in Figs.~\ref{fig:low}, \ref{fig:high} and \ref{fig:all} the weight averaged rotation angle over the whole beam profile. 

Our method differs from (and is more robust than) conventional Faraday rotation measurements: Typically, the rotation angle is derived from differential measurement of light in two orthogonal linear polarisations $\theta = 1/2\arcsin S_1/S_0$, which should ideally be situated at $\pm 45 ^\circ$ from the input polarisation and are most sensitive to small rotation angles. This method is only valid for negligible beam ellipticity.

Our tomographical method allows to identify any polarisation state without bias for a range of different input polarisations. Being able to reconstruct the full spatial polarisation structure of the beam after interaction with the atoms not only allows us to calculate the exact rotation, but also to investigate the effect on spatial modes. 

\section{data}

\begin{figure}
    \centering    \includegraphics[width=\linewidth]{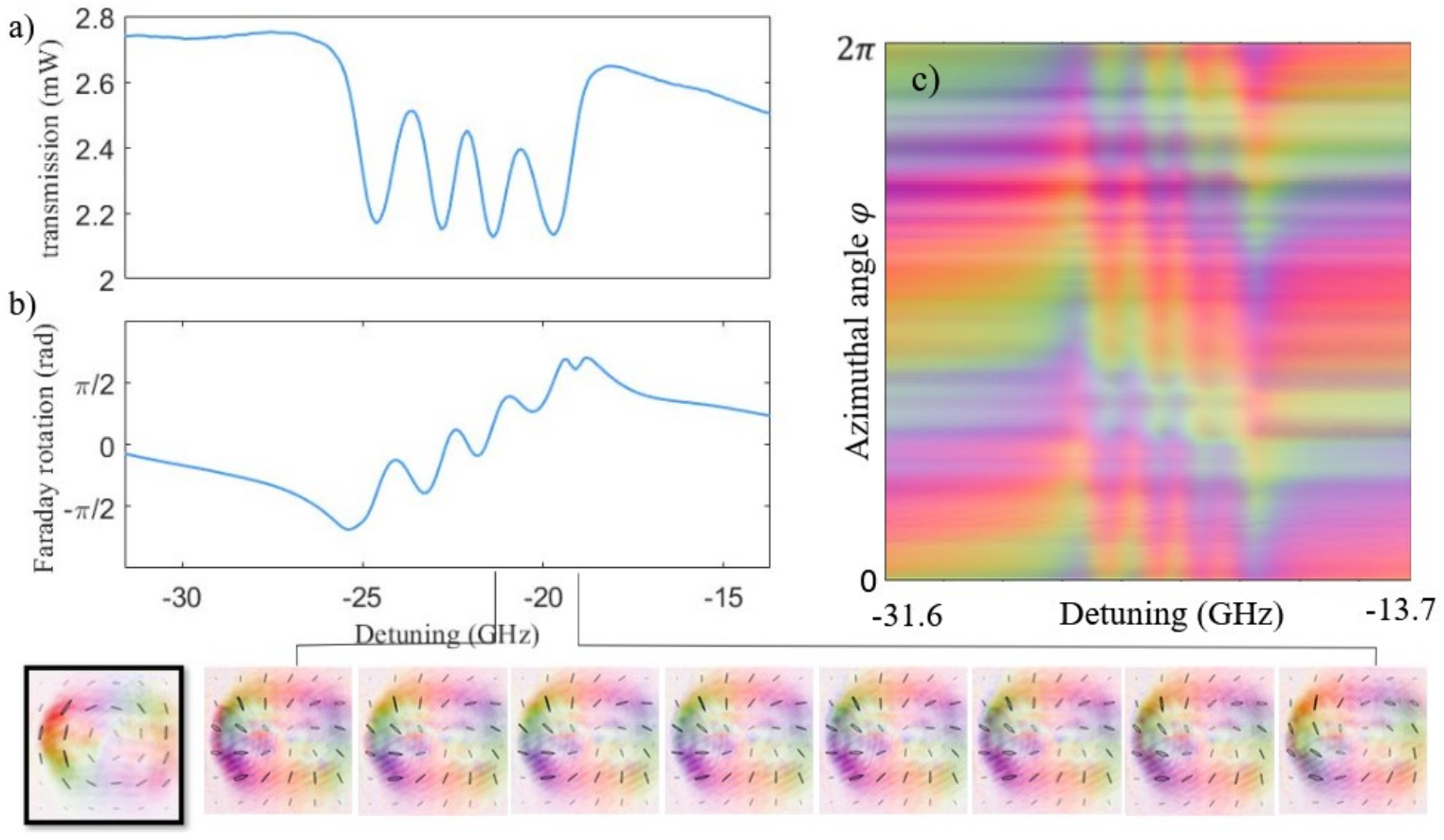}
    \caption{Frequency dependent absorption (a) and Faraday rotation (b) at a temperature of 92.1~$^\circ$C. The bottom line shows the azimuthal input beam (framed in black) as well as a series of polarisation profiles after Faraday rotation at evenly spaced detunings from -21.39~GHz and -19.71~GHz, spanning rotation angles between 0.5 and 1.6 radians. (c) Map of the local polarisations around the unwrapped probe beam using the colour scheme of Fig.~\ref{fig:setup}(b).}
    \label{fig:low}
\end{figure}

\begin{figure}
    \centering    \includegraphics[width=\linewidth]{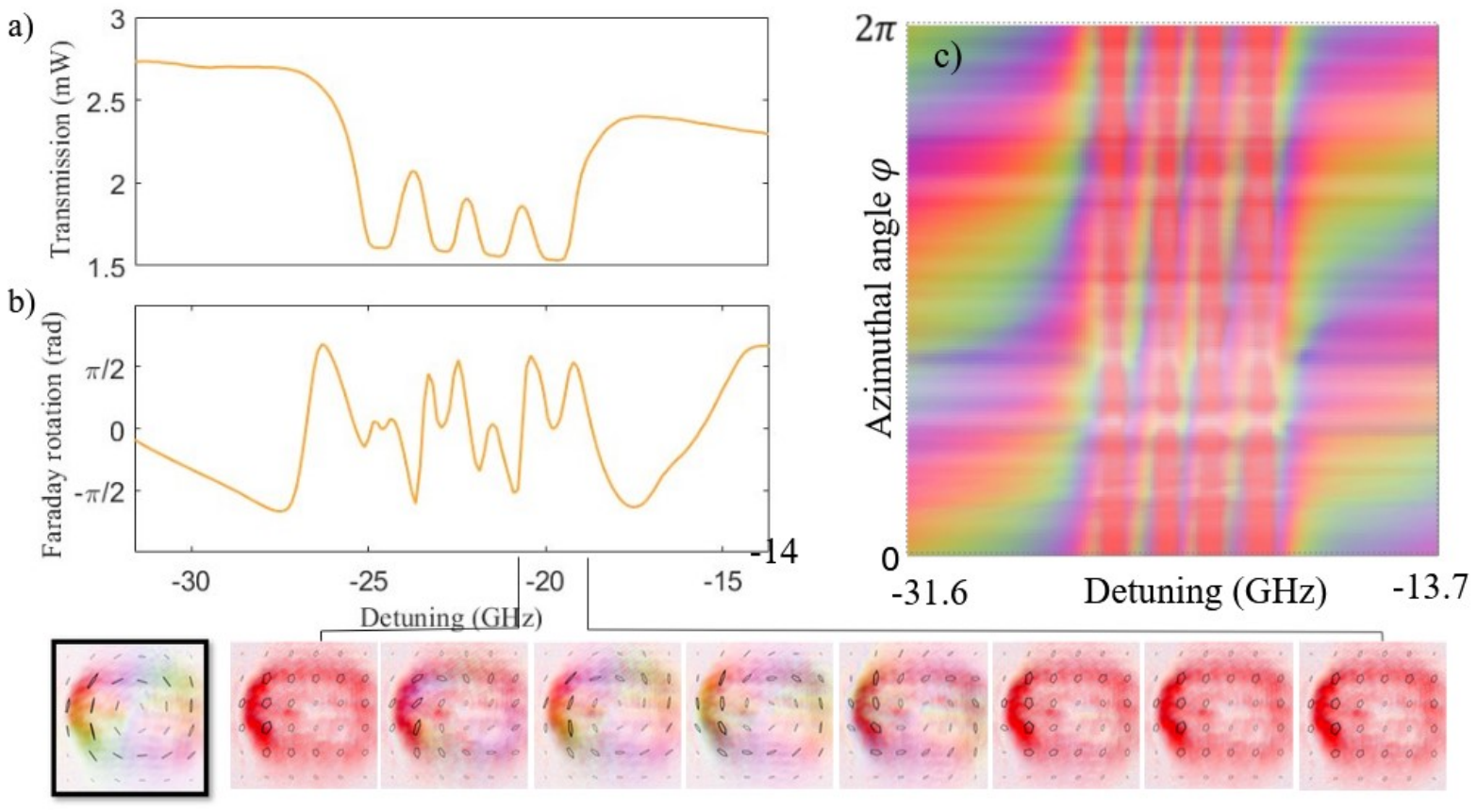}
    \caption{Same as Figure \ref{fig:low} but for a temperature of 114.9~$^\circ$C. The bottom row shows that for higher atomic densities the left circular component of the azimuthal light is largely absorbed close to resonance, leading to predominantly right circular polarised probe light, and an apparent counterrotation as the frequency is scanned over the resonance line. The effective rotation still ranges from -1.6 to 1.8 radians.}
    \label{fig:high}
\end{figure}

\begin{figure}
    \centering    \includegraphics[width=\linewidth]{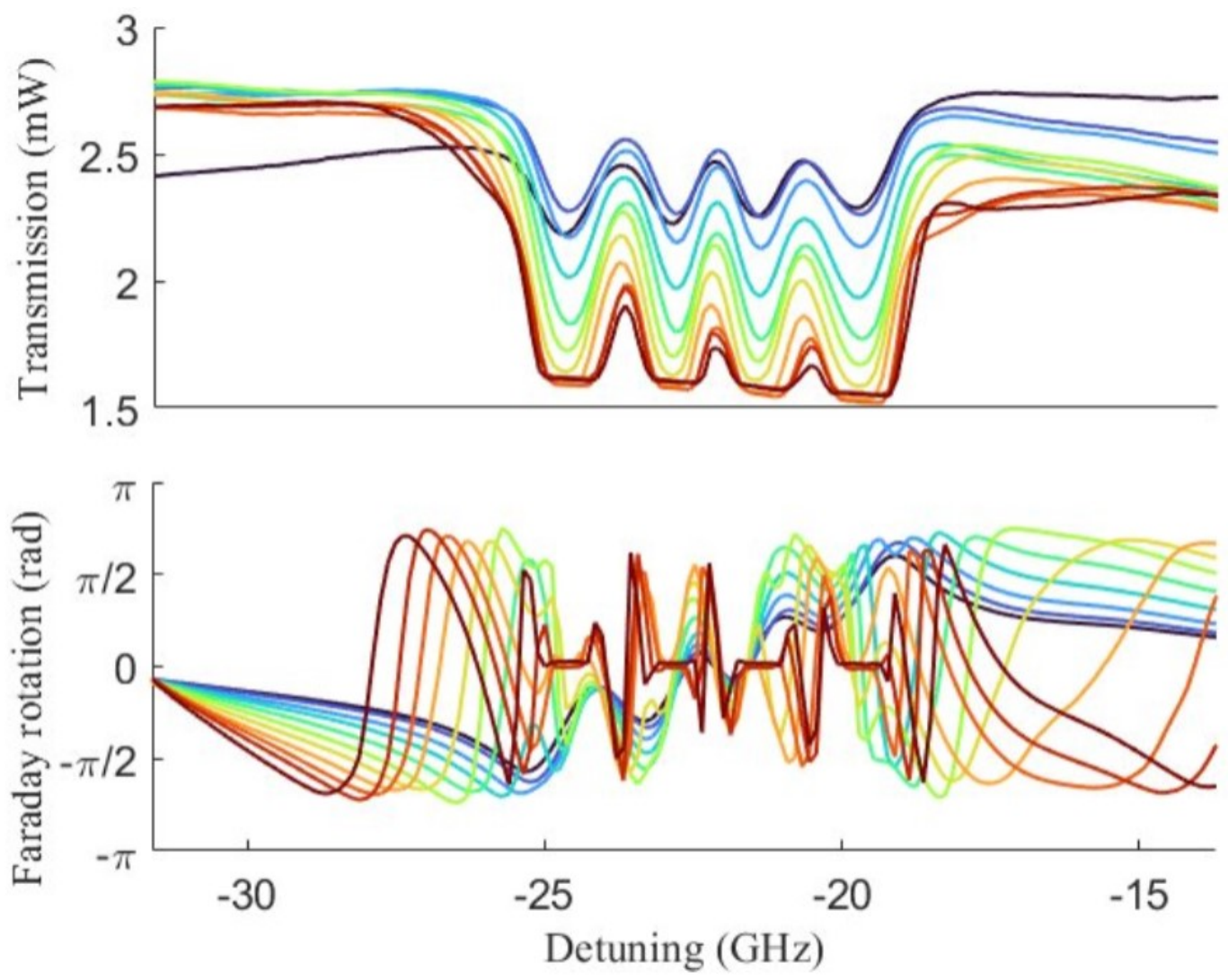}
    \caption{Absorption and Farady rotation across a range of temperatures, showing the increasing absorption and saturation of the atomic transitions, as well as large fluctuations in Faraday rotation.}
    \label{fig:all}
\end{figure}

Subtracting the local polarisation orientation of each pixel from the theoretical azimuthal angle (orthogonal to the azimuthal polar coordinate) yields the total Faraday rotation for each frame. In this experiment, this was done pixel by pixel, with the total rotation being calculated via a weighted average, taking into account local intensity in order to reduce background noise.

Figure~\ref{fig:low} summarises our measurements for a relatively low atomic density (for a temperature identified via {\it Elecsus} as 92.1 $^\circ$C). The transmission lines are not saturated (Figure~\ref{fig:low}(a)), and the Faraday rotation reveals an `envelope' of dispersion curves, corresponding to the composite dispersion curve of each of the individual atomic lines (Figure~\ref{fig:low}(a)). Transmission and Faraday angle are qualitatively in good agreement with the theoretical plots shown in Fig.~\ref{fig:theory}, but we note that the measured rotation angle significantly exceeds the predicted angle, spanning a range of more that $\pi$. While the spectral shape is fixed by the Zeeman splitting, the atomic density is underestimated in the weak probe simulation, leading to the quantitative difference in rotation angle.  When the Faraday rotation angle is $\pm \pi/2$, the azimuthal input beam is transformed into a radial beam, and for other rotation angles the polarisation state is a superposition of radial and azimuthal with a twisted structure. 
Another way to visualize this is shown in Figure~\ref{fig:low}(c). Here, the beam is ``unwrapped" by mapping the azimuthal beam angle (averaged over the high intensity ring) onto the vertical axis and the detuning on the horizontal axis. The polarisation colour map visualises the rotation of the local polarisation within the beam. 

Figure \ref{fig:high} shows the same curves at a higher atomic density temperature (for a temperature identified via {\it Elecsus} as 114.9 $^\circ$C). Here, the absorption is strong enough to significantly alter the ellipticity of the beam. If the absorption is saturated, all $\sigma_+$ polarised light is removed from the beam, and only $\sigma_-$ remains (indicated in our colour map as red). Circular dichroism means that Faraday rotation becomes meaningless (and unmeasurable) close to resonance as the linear polarised beam is turned circular at high atomic densities. The rotation for detunings close to resonance are effectively reduced, so that additional ``dips" appear in the rotation curve, corresponding to frequency ranges over which the polarisation rotates backwards. Figure \ref{fig:high}(c) highlights that the beam now undergoes a full rotation between each absorption dip.

Figure \ref{fig:all} shows transmission and rotation angles for a range of atomic densities (with estimated temperatures between 79$^\circ$C and 132$^\circ$C). As the temperature increases, two things happen: the Faraday rotation angle begins to approach and eventually exceed 2$\pi$, and the absorption begins to saturate. When the transition is saturated, all right circularly polarised light in the beam is absorbed, and all that is left is left circularly polarised, i.e. an equal superposition of radial and azimuthal. In this case, a rotation angle can no longer be identified. For temperatures upward of $\approx $110$^\circ$C, the beam undergoes a full Faraday rotation between any two absorption dips. Data taken at 114.9$^\circ$C is shown in more detail in figure \ref{fig:high}, with full polarisation plots for a series of evenly spaced individual frames between -21.39~GHz and -19.71~GHz. As mentioned previously, the figure appears to jump when the rotation angle crosses $\pi$. We can observe how these jumps move further from the absorption dips for higher temperatures, indicating a steeper slope, indicating stronger rotation. 
An additional effect in the measurement at these temperatures is that polarisation only measures orientation, and as such repeats at $\pi$ rather than $2\pi$. Therefore, the figure appears to jump at values when the curve crosses $\pi$. 

The fact that we focused the light highlights the spatial resolution achieved by these systems. Azimuthally polarised light is unique in that it does not generate 3D polarisation when focused, instead preserving its initial structure and achieving smaller spot sizes. These vector beams also contain every linear polarisation, which highlights the circular eigenmodes and the universality of the system. 

\section{conclusion}

In this paper we have illistrated how the Faraday effect can transform vector beams to their orthogonal structures near resonance in the HPB regime.
As the absorption dips are polarisation dependent and occur in clusters of four, the refractive indeces for right and left circularly polarised light vary quite drastically between individual absorption lines. As such, local polarisations are rotated and the beam becomes a superposition of radial and azimuthal. 

Raising the temperature of the system increases the number density of the atoms, deepening the absorptions. Once the absorption saturates, the $\sigma_-$ component of the light is entirely absorbed. Between these dips, the light now undergoes a full Faraday rotation, transforming from azimuthal to radial and back multiple times over the course of a scan. 

We measured the Faraday rotation via full spatially resolved Stokes tomography, an overcomplete measurement which sits in contrast with the more traditional approach, which projects into a single basis and measures with photodiodes. This allows us to measure for higher order properties such as the spatial structure and remains valid for circular dichroism. 

This system is highly sensitive to detunings and can be controlled via the magnetic field and the temperature. Additionally, such transformations should also be generalisable to other types of vector beams.

\section*{Author Contributions}
\textbf{Sphinx J. Svensson}: Data curation, Visualization, Writing – original draft, Software, Conceptualisation. \textbf{Craig J. A. Millar}: Data curation. \textbf{Danielle Pizzey}: Development of the original setup. \textbf{Ifan G. Hughes}: Conceptualization, Resources. \textbf{Sonja~Franke-Arnold}: Conceptualization, Supervision, Visualization, Writing – original draft, Funding acquisition, Resources. All authors contributed to reviewing \& editing the manuscript.

\bibliography{bib}

\begin{acknowledgments}
We are grateful for Dr Steven Wrathmall for providing the ElecSUS support and generating the data shown in Figure \ref{fig:theory}. This work was funded, in part, by the QuantERA II Programme with funding received via the EU H2020 research and innovation programme under Grant No. 101017733, EPSRC under Grant No. EP/Z000513/1 (V-MAG). 
\end{acknowledgments}

\section*{disclosures} 
The authors declare no conflict of interest.

\end{document}